\documentclass[conference]{IEEEtran}
\usepackage{adjustbox}
\usepackage{subfig}
\usepackage{graphicx}
\usepackage{textcomp}
\usepackage{xcolor}
\usepackage{soul}
\usepackage{url}
\newtheorem{definition}{Definition}

\newtheorem{theorem}{Theorem}

\usepackage{booktabs}
\usepackage{algorithm}
\usepackage{algorithmic}
\usepackage{colortbl} 
\usepackage{arydshln} 
\usepackage{bbding}
\usepackage{amsmath}
\usepackage{amssymb}
\usepackage{multirow} 
\usepackage{multicol}
\usepackage[utf8]{inputenc}
\AtBeginDocument{%
  \providecommand\BibTeX{{%
    \normalfont B\kern-0.5em{\scshape i\kern-0.25em b}\kern-0.8em\TeX}}}

\usepackage{amssymb}

\begin{document}

\title{When FinTech Meets Privacy: Securing Financial LLMs with Differential Private Fine-Tuning}


\author{
    \IEEEauthorblockN{
        Sichen Zhu\IEEEauthorrefmark{1}, 
        Hoyeung Leung\IEEEauthorrefmark{1}, 
        Xiaoyi Wang\IEEEauthorrefmark{2}, 
        Jia Wei\IEEEauthorrefmark{3}, 
        Honghui Xu\IEEEauthorrefmark{3}\IEEEauthorrefmark{4}
    }
    \IEEEauthorblockA{\IEEEauthorrefmark{1}Georgia Institute of Technology, Atlanta, GA, USA\\}
    \IEEEauthorblockA{\IEEEauthorrefmark{2}Sichuan University of Media and Communications, Chengdu, Sichuan, China\\}
    \IEEEauthorblockA{\IEEEauthorrefmark{3}Kennesaw State University, Marietta, GA, USA\\}
    \IEEEauthorblockA{\IEEEauthorrefmark{4}Corresponding author: Email: hxu10@kennesaw.edu}
}


\maketitle






\begin{abstract}

The integration of Large Language Models (LLMs) into financial technology (FinTech) has revolutionized the analysis and processing of complex financial data, driving advancements in real-time decision-making and analytics. 
With the growing trend of deploying AI models on edge devices for financial applications, ensuring the privacy of sensitive financial data has become a significant challenge.
To address this, we propose DPFinLLM, a privacy-enhanced, lightweight LLM specifically designed for on-device financial applications. DPFinLLM combines a robust differential privacy mechanism with a streamlined architecture inspired by state-of-the-art models, enabling secure and efficient processing of financial data. 
This proposed DPFinLLM can not only safeguard user data from privacy breaches but also ensure high performance across diverse financial tasks.
Extensive experiments on multiple financial sentiment datasets validate the effectiveness of DPFinLLM, demonstrating its ability to achieve performance comparable to fully fine-tuned models, even under strict privacy constraints.

\end{abstract}







\section{Introduction}\label{sec:introduction}

The proliferation of LLMs has revolutionized natural language understanding and generation, driving significant advancements in the FinTech sector. These models excel in processing complex financial data, enabling applications such as sentiment analysis~\cite{zhang2023enhancing}, risk management~\cite{yang2024financial}, fraud detection~\cite{boulieris2024fraud}, and credit scoring~\cite{sanz2024credit}. By providing actionable insights and facilitating real-time decision-making, LLMs have become indispensable tools in modern financial services, empowering FinTech solutions to deliver smarter, faster, and more secure financial operations~\cite{xue2024domain}.

Recently, there has been a growing trend toward deploying AI models on edge devices for financial applications~\cite{hassan2024applications}. 
In light of this, on-device financial LLMs will offer several advantages, including real-time data processing, reduced dependency on cloud services, and enhanced user data privacy—critical components in the evolving FinTech landscape. However, this trend also introduces significant challenges. On-device models~\cite{xu2024device,chen2024octo,chen2024octopusv2,chen2024octopusv3,chen2024octopusv4} must achieve a delicate balance between computational efficiency, privacy protection, and task performance, making their design and training particularly complex. Addressing these challenges is essential to advancing the integration of LLMs into secure, efficient, and innovative FinTech ecosystems.


The sensitive nature of financial data amplifies the challenges associated with deploying on-device financial LLMs. Membership inference attacks~\cite{zhang2024no} and model inversion attacks~\cite{nayan2024sok} on these AI models pose significant risks, enabling unauthorized access to private financial information. While existing differential privacy techniques have been explored across three key phases, input dataset preparation~\cite{martin2024artificial}, model training~\cite{liu2024dpdr}, and model output generation~\cite{zhang2024no}, their application to on-device financial LLMs remains under-researched. Bridging this gap is crucial to ensuring the secure and effective deployment of financial LLMs on edge devices, safeguarding sensitive data while maintaining financial LLMs' performance.

To address these challenges, we introduce DPFinLLM, a novel on-device financial large language model that integrates a lightweight architectural design with a robust differential privacy mechanism. DPFinLLM employs a privacy-enhanced training pipeline to safeguard sensitive financial data while maintaining high performance across various financial tasks. Drawing inspiration from state-of-the-art models like Llama2 and ChatGLM2, DPFinLLM features a streamlined architecture optimized for edge devices. By incorporating Low-Rank Adaptation (LoRA) for fine-tuning, the model achieves computational efficiency and supports task-specific optimization with minimal resource requirements. The key contributions of this paper are summarized as follows:

\begin{itemize}
    \item We propose DPFinLLM, a privacy-enhanced on-device financial LLM that integrates differential privacy techniques with a lightweight architectural design tailored for edge devices.
    \item A robust differential privacy mechanism is incorporated into the fine-tuning process to protect sensitive financial data from privacy breaches.
    \item Comprehensive experiments on multiple financial sentiment datasets validate the effectiveness of DPFinLLM, demonstrating performance comparable to baseline models even under strict privacy constraints.
\end{itemize}

The remainder of this paper is structured as follows: Section~\ref{sec:related_works} reviews existing research on financial LLMs and privacy-preserving techniques. Section~\ref{sec:approach} details the architectural design and privacy-preserving training framework of DPFinLLM. Section~\ref{sec:experiments} presents the experimental setup and results, and Section~\ref{sec:conclusion} concludes with insights and directions for future research.

\section{Related Works}\label{sec:related_works}

This section will conclude the related work of financial language models and review the current mainstream privacy-preserving learning approaches. 

\subsection{Financial Large Language Models}\label{subsec:FLLM}

Recently, financial language models have demonstrated remarkable capabilities in handling complex tasks in the financial sector, including sentiment analysis~\cite{delgadillo2024finsosent}, risk management~\cite{yang2024financial}, financial fraud detection~\cite{boulieris2024fraud}, and credit scoring~\cite{sanz2024credit}. 
By leveraging their predictive power, users in the economic domain can make more informed decisions, enabling better planning and strategic execution.
Modern fine-tuned LLMs, such as FinGPT~\cite{yang2023fingpt} and FinBERT~\cite{huang2023finbert}, have been specifically developed to handle natural language processing tasks in financial datasets, showcasing exceptional domain-specific adaptability. These models excel at processing both structured and unstructured data~\cite{li2023extracting} from various sources, including APIs, web scraping tools, and direct database access. Their ability to integrate real-time data streams into model training provides outputs that reflect the most current market conditions or issue statuses, offering users actionable insights in a dynamic financial landscape~\cite{zhao2024revolutionizing}.
As financial language models continue to evolve, a notable trend is emerging toward the development of on-device financial LLMs.
However, designing on-device financial LLMs presents considerable challenges. Unlike cloud-based models, which can leverage extensive computational resources, on-device models must operate within the constraints of limited processing power, memory, and energy. 
Overcoming these hurdles will be crucial to unlocking the full potential of on-device financial LLMs and transforming the way financial insights are generated and utilized.

\subsection{Privacy-Preserving Learning Mechanisms}\label{subsec:PPL}

Privacy-preserving approaches address critical AI cybersecurity challenges by implementing mechanisms across three key phases: input dataset preparation, model training, and model output generation. 
(1) In the input dataset preparation phase, privacy is safeguarded by injecting noise (e.g., Gaussian or Laplace) into dataset features~\cite{majeed2023ai,martin2024artificial}, replacing sensitive text with anonymous tokens or artificial labels, and encrypting data to minimize leakage risks~\cite{yang2024ai}. 
(2) During the model training process, some techniques ensure privacy by clipping gradients to limit individual data point influence and adding Gaussian noise to gradients for differential privacy~\cite{liu2024dpdr,fu2022sa}.
(3) In the model output generation phase, methods introduce noise to prompts or outputs to protect sensitive information, followed by coherence refinement to maintain utility. Additionally, output filtering can replace sensitive terms with anonymized equivalents to prevent disclosure~\cite{zhang2024no}.
While these strategies effectively address privacy concerns in traditional large language models, there is a notable lack of research investigating data privacy mechanisms specifically for on-device financial LLMs. Developing privacy-preserving solutions tailored for on-device models is critical, as such systems face unique challenges, including constrained computational resources and heightened sensitivity to data privacy breaches.

In this paper, we propose an on-device differential privacy-enhanced financial LLM (called DPFinLLM), tackling two critical challenges: designing a lightweight architecture suitable for resource-constrained edge devices and incorporating a differential privacy mechanism during model training to protect users’ sensitive data effectively.


\section{DPFinLLM}\label{sec:approach}

The proposed model, DPFinLLM, is an on-device differential privacy-enhanced financial large language model designed to address two primary challenges: creating a lightweight architecture for deployment on resource-constrained edge devices and integrating a robust differential privacy mechanism to safeguard sensitive financial data. DPFinLLM employs a transformer-based architecture with modifications inspired by Llama2.
For fine-tuning, the model leverages LoRA, a parameter-efficient technique that reparameterizes weight updates during training, significantly reducing memory and computational costs. To ensure privacy preservation, the model incorporates an ($\epsilon, \delta$)-differential privacy framework during training. 
The framework limits the influence of individual samples on parameter updates using gradient clipping and adds Gaussian noise to batch gradients.
By fine-tuning DPFinLLM with a well-structured loss function and optimizing hyperparameters such as privacy leakage bounds ($\epsilon, \delta$), gradient norm limits, and batch sizes, the model ensures robust privacy protection while maintaining high performance for financial-specific tasks. This architecture and training approach make DPFinLLM suitable for secure and efficient deployment in sensitive financial environments.

\subsection{Lightweight Financial LLM Design and Fine-Tuning}\label{subsec:Fine_LLM_LoRA}

We denote the basic LLM learning process as $\pi_{\theta} \in \Pi$, parameterized by $\theta \in \Theta$ through a neural network. The input prompt $\mathbf{x}$ is tokenized into tokens from a predetermined vocabulary $\mathcal{V}$, represented as $\mathbf{x} = [x_1, …, x_n], x_i \in \mathcal{V}$. The output sequence is denoted as $\mathbf{y}$. The generated outputs $\mathbf{y}$, given input $\mathbf{x}$, are expressed as:
\begin{equation}
\label{eq:auto_x_y}
\pi_{\theta} (\mathbf{y} | \mathbf{x} ) = \prod_{t = 1}^{T} \pi \left( y_t | x_1, x_2, …, x_n, ,, y_1, y_2, …, y_{t-1} \right),
\end{equation}
where $y_i$ is sampled from the conditional probability distribution $\pi_{\theta} (\cdot | \mathbf{x})$, and the next token $y_{t}$ is generated based on all previously generated tokens $(y_1, y_2, …, y_{t-1})$.

We utilize a transformer-based LLM, where the core component is the attention function, mapping a query and a set of key-value pairs to an output:
\begin{equation}
\label{eq:transformer}
\text{Attention} (Q, K, V) = \text{softmax} \left ( \frac{QK^T}{\sqrt{d_k}} \right ) V,
\end{equation}
where $Q$, $K$, and $V$ represent the query, key, and value matrices, respectively, and $d_k$ is the size of the hidden dimensions. The factor $1/\sqrt{d_k}$ serves as a scaling term. To capture semantic and contextual information from input tokens with varying emphases at different positions, we employ a multi-head attention mechanism~\cite{vaswani2017attention}:
\begin{equation}
\label{eq:attn}
\text{MultiHead} (Q, K, V) = \text{Concat} (\text{head}_1, …, \text{head}_h) W^O,
\end{equation}
where $\text{head}_i = \text{Attention} \left (QW_i^Q, KW_i^K, VW_i^V \right )$, with parameter matrices $W_i^{Q}$, $W_i^{K}$, $W_i^{V}$, and $W^{O}$.

To pursue a lightweight architecture for the LLM, we adopt a modified attention mechanism inspired by Llama2~\cite{touvron2023llama}, replacing traditional multi-head attention to improve performance and reduce memory costs as context window and batch sizes increase in larger models. Grouped-query attention~\cite{ainslie2023gqa} enables the sharing of key and value projection matrices ($W_i^K$, $W_i^V$) across multiple heads without compromising model performance. Furthermore, we incorporate RMSNorm~\cite{zhang2019root} for layer normalization, with the activation function SwiGLU~\cite{shazeer2020glu}, to mitigate the internal covariate shift issue often encountered in vanilla neural networks.

Though trained on generic datasets, a lightweight architecture for financial LLMs can effectively adapt to specific domains by being fine-tuned on internet-scale financial data from diverse tasks. This adaptation is enabled by leveraging the basic LLM architecture. For positional embeddings, we employ rotary positional embeddings~\cite{su2024roformer}, an interpretable method that integrates relative positional information into the rotation of context representations, enhancing the model’s ability to capture sequential dependencies.

For the fine-tuning of our lightweight LLM, we utilize the Low-Rank Adaptation (LoRA)~\cite{hu2021lora} method alongside instruction tuning. LoRA is a parameter-efficient approach for fine-tuning pre-trained LLMs to specific tasks. It operates under the assumption that updates to model weights during fine-tuning have a low “intrinsic rank.” For a pre-trained weight matrix $W_0 \in \mathbb{R}^{d \times k}$, the weight update during fine-tuning is reparameterized as:
\begin{equation}
\label{eq:LoRA}
W_0 + \Delta W = W_0 + BA,
\end{equation}
where $B \in \mathbb{R}^{d \times r}$ and $A \in \mathbb{R}^{r \times k}$. The rank $r$ of $\Delta W$ controls the number of trainable parameters and is typically small, $r \ll \min(d, k)$. During fine-tuning, only $A$ and $B$ are updated, while the pre-trained weight $W_0$ remains frozen. For an input vector $x$ and an output vector $h$, the forward pass is computed as:
\begin{equation}
\label{eq:forward}
h = (W_0 + \Delta W )x = W_0x + \Delta W x = W_0 x + BAx,
\end{equation}
where $\Delta W x$ is further scaled by $\alpha / r$, allowing the rank of $\Delta W$ to remain intact while scaling the impact of $\Delta W$ without increasing the number of training parameters. LoRA significantly reduces memory and computational costs by avoiding expensive matrix multiplications, thereby enabling the fine-tuning of LLMs in a computation- and parameter-efficient manner.

\subsection{DP Mechanism for Secure Financial Model Training}\label{subsec:DPL}

To protect sensitive financial data, we propose training the financial model using an ($\epsilon, \delta$)-differential privacy (DP) mechanism~\cite{dwork2006our}. Specifically, in this ($\epsilon, \delta$)-DP training framework, the influence of individual samples on parameter updates is limited during training. For training samples ${x_1, …, x_N}$, the loss function with respect to $\theta$, minimized during neural network training, is expressed as:
\begin{equation}
\label{eq:loss}
\mathcal{L}(\theta) = \frac{1}{N} \sum_i \mathcal{L} (\theta, x_i).
\end{equation}

For a random batch of training samples $\mathcal{B}$, the gradient for each individual sample $x_i \in \mathcal{B}$ at training step $t$ is computed as $\mathbf{g}t(x_i) = \nabla{\theta_t} \mathcal{L} (\theta_t, x_i)$. To constrain the influence of any single sample, the gradient is clipped as follows:
\begin{equation}
\label{eq:gradient}
\overline{\mathbf{g}}_t(x_i) = \frac{\mathbf{g}_t (x_i)}{\max \left( 1, , \frac{||\mathbf{g}_t(x_i)||_2}{C} \right )},
\end{equation}
where $C$ is the gradient norm bound, preventing the model from being overly influenced by any single training sample. To further protect against memorization of training data, Gaussian noise is added to the average gradient over the batch $\mathcal{B}$:
\begin{equation}
\label{eq:average_grad}
\Tilde{\mathbf{g}}_t = \frac{1}{L} \left ( \sum_{i=1}^{L} \overline{\mathbf{g}}_t (x_i) + \mathcal{N}\left (0, \sigma^2 C^2 \mathbf{I} \right )\right ),
\end{equation}
where $\sigma$ is the noise scale, and $L$ is the lot size—the number of samples whose gradients are computed and averaged. To limit memory consumption, the batch size $\mathcal{B}$ is typically much smaller than the lot size $L$, as gradient clipping and noising are performed per sample. Finally, a standard gradient descent step updates $\theta$ as:
\begin{equation}
\label{eq:updates}
\theta_{t+1} = \theta_t - \eta_t \Tilde{\mathbf{g}}_t,
\end{equation}
where $\eta_t$ is the learning rate.

From an engineering perspective, a larger batch size can improve convergence during training. However, increasing the batch size also raises the privacy cost, as reflected in Theorem~\ref{thm:1}, derived from Definition~\ref{def:1}.
\begin{definition}
\label{def:1}
A randomized mechanism $\mathcal{M}: \mathcal{D} \rightarrow \mathcal{R}$, with domain $\mathcal{D}$ and range $\mathcal{R}$, satisfies ($\epsilon, \delta$)-differential privacy~\cite{abadi2016deep} if, for any two adjacent inputs $d, d’ \in \mathcal{D}$ and any subset of outputs $S \subseteq \mathcal{R}$, it holds that:
\begin{equation}
\label{eq:differential_privacy}
\Pr[\mathcal{M}(d)] \leq e^{\epsilon} , \Pr [\mathcal{M} (d’)] + \delta,
\end{equation}
where $\mathcal{M}(d), \mathcal{M}(d’) \in S$.
\end{definition}
\begin{theorem}
\label{thm:1}
There exist constants $c_1$ and $c_2$ such that, given the lot size $L$, sample size $N$, sampling probability $q = L / N$, and number of steps $T$, for any $\epsilon < c_1 q^2 T$, the DPSGD algorithm satisfies ($\epsilon, \delta$)-DP for any $\delta > 0$, provided
\begin{equation}
\label{eq:thm1}
\sigma \geq c_2\frac{q \sqrt{T \log(1/\delta)}}{\epsilon}.
\end{equation}
\end{theorem}

This theorem establishes that the algorithm is ($\epsilon, \delta$)-DP, meaning the outputs are perturbed by a factor governed by $e^{\epsilon}$, ensuring indistinguishability of datasets differing by a single data point. The additive term $\delta$ slightly relaxes the constraint and is typically chosen to be smaller than $1/|d|$. The parameters $\epsilon$ and $\delta$ quantify \textit{privacy leakage}~\cite{li2021large}, mathematically defining the privacy-preserving goal. Smaller values of $\epsilon$ and $\delta$ guarantee that differences in the input dataset minimally affect the algorithm’s random output, reducing the risk of private data exposure to inference attacks.

In conclusion, we will apply the proposed differential privacy learning algorithm to fine-tune our financial LLM, as detailed in Section~\ref{subsec:Fine_LLM_LoRA}, using the loss function in Eq.~\eqref{eq:loss} and exploring various hyperparameter settings, including $\epsilon$, $\delta$, gradient norm bounds, and batch sizes.

\section{Experiments}\label{sec:experiments}

This section outlines the experimental setup and presents a thorough analysis of the results, highlighting the effectiveness of our proposed DPFinLLM.
The open-source codes of the experiments can be found in \url{https://github.com/SichenZhu/DP_FinLLM}.

\subsection{Experimental Settings}\label{subsec:experimental_setting}
The datasets, training setup, hyperparameter settings, baselines, and performance metrics are described below.

\subsubsection{Datasets}\label{subsubsec:dataset}

We evaluate the effectiveness of our proposed DPFinLLM model using four sentiment analysis datasets from the financial domain. 
(1) The first dataset, FPB, contains 4,846 news entries that cover a diverse range of small and large companies, various industries, and multiple news sources. The sentiment labels in this dataset are assigned from an investor’s perspective~\cite{malo2014good}. 
(2) The second dataset, FIQA, includes 1,213 entries sourced from financial news headlines and microblogs, with annotations for target entities, sentiment scores, and aspects~\cite{maia201818}.
(3) The third dataset, TFNS, comprises 11,931 finance-related tweets collected via the Twitter API, capturing social media sentiment~\cite{tfns}. 
(4) Lastly, NWGI is composed of 20,231 entries generated using GPT-based instructions, providing a wide range of sentiment-rich content~\cite{nwgi}. 
The datasets are organized in the following way for fine-tuning of LLMs:

\textbf{Instruction}: ``What is the sentiment of this news/tweet? 
Please choose an answer from {negative/neutral/positive}''

\textbf{Input}: [input] \textbf{Answer}: [output]

\subsubsection{Baseline}\label{subsubsec:baseline}

We use two models as baselines for performance comparison. 
(1) Llama2~\cite{touvron2023llama}, is an open-source large language model (LLM) developed by Meta. Llama2 is a highly capable chat model that surpassed existing open-source chat models on most benchmarks at the time of its release. The state-of-the-art performance of FinGPT in sentiment analysis was achieved using the Llama2 model family. 
(2) ChatGLM2~\cite{glm2024chatglm}, is an open-source bilingual general language model. ChatGLM2 served as the base model for Financial GPT, which also achieved SOTA performance in sentiment analysis. 
We compare the performance of our proposed DPFinLLM against the original Llama2-7B and ChatGLM2-6B models, both trained on a financial multi-task dataset that includes all four sentiment analysis datasets.

\subsubsection{Performance Metrics}\label{subsubsec:metrics}

The performance evaluation metric is accuracy and three different calculations for F1-score ``micro'', ``macro'' and ``weighted''). The micro F1 score counts total true positives, false negatives and false positives. The macro F1 score averages the F1 score calculated for each label. The weighted F1 score takes the weighted average F1 score from each label to take the label imbalanceness into consideration.

\subsubsection{Training Setup}\label{subsubsec:architecture}

We adopt the same network architecture design as Llama2 and ChatGLM2 while incorporating the proposed DPFinLLM framework during the training process of the LLM, as outlined in Section~\ref{sec:approach}. Specifically, the base model is fine-tuned using the LoRA method described in Section~\ref{subsec:Fine_LLM_LoRA}, with the differential privacy (DP) mechanism detailed in Section~\ref{subsec:DPL} integrated into the fine-tuning process. This approach is tailored to address the sentiment analysis task, ensuring both performance optimization and robust privacy protection.

\subsubsection{Hyperparameter Settings}\label{subsubsec:parameters}

We experiment with various hyperparameter settings during training and testing to optimize model performance. 
Notably, for certain experiments involving the differential privacy mechanism, the parameter $\delta$ is set to $1 / |d|$, where $d$ represents the size of the training dataset.

\begin{table}[htbp]
\caption{Comparison Performance of Sentiment Analysis on FPB Dataset (Baselines v.s. Our DPFinLLM)}
\label{tab:fpb_results}
\renewcommand{\arraystretch}{1.3}
\begin{center}
\resizebox{\linewidth}{!}{\begin{tabular}{c||c|c|c|c}
\hline\hline
\multicolumn{1}{c||}{\bf Metric} &\multicolumn{1}{c|}{\bf Accuracy} &\multicolumn{1}{c|}{\bf F1 macro} &\multicolumn{1}{c|}{\bf F1 micro}  &\multicolumn{1}{c}{\bf F1 weighted } \\ 
\hline
Llama2-7B (Llama2-based) &0.46947 &0.52394 &0.46947 &0.40989\\
FinGPT-llama2-mt (Llama2-based) &0.79785 &0.75539 &0.79785 &0.77654 \\
FinGPT v3.2 (Llama2-based) &0.86634 &0.85190 &0.86634 &0.86371 \\
\textbf{Llama2-based DPFinLLM with DP $\epsilon = 8.0$} & \textbf{0.79785} & \textbf{0.77880} &\textbf{0.79785} & \textbf{0.79524} \\
\hline
ChatGLM2-6B (ChatGLM2-based) &0.45462 &0.47733 &0.45462 &0.36403 \\
FinGPT v3.1 (ChatGLM2-based) &0.85231 &0.83502 &0.85231 &0.85100 \\
\textbf{ChatGLM2-based DPFinLLM with DP $\epsilon = 8.0$} &\textbf{0.47030} &\textbf{0.49355} &\textbf{0.47030} & \textbf{0.39147}\\
\hline\hline
\end{tabular}}
\end{center}
\end{table}

\begin{table}[htbp]
\caption{Comparison Performance of Sentiment Analysis on FIQA Dataset (Baselines v.s. Our DPFinLLM)}
\label{tab:fiqa_results}
\renewcommand{\arraystretch}{1.3}
\begin{center}
\resizebox{\linewidth}{!}{\begin{tabular}{c||c|c|c|c}
\hline\hline
\multicolumn{1}{c||}{\bf Metric} &\multicolumn{1}{c|}{\bf Accuracy} &\multicolumn{1}{c|}{\bf F1 macro} &\multicolumn{1}{c|}{\bf F1 micro}  &\multicolumn{1}{c}{\bf F1 weighted } \\ 
\hline 
Llama2-7B (Llama2-based) &0.78909 &0.59376 &0.78909 &0.77448 \\
FinGPT-llama2-mt (Llama2-based) &0.43636 &0.45265 &0.43636 &0.52937 \\
FinGPT v3.2 (Llama2-based) &0.75273 &0.67581 &0.75273 &0.80064 \\
\textbf{Llama2-based DPFinLLM with DP $\epsilon = 8.0$} &\textbf{0.80727} &\textbf{0.61251} &\textbf{0.80727} &\textbf{0.78646}\\
\hline
ChatGLM2-6B (ChatGLM2-based) &0.83636 &0.57016 &0.83636 &0.80312 \\
FinGPT v3.1 (ChatGLM2-based) &0.82909 &0.73848 &0.82909 &0.84298 \\
\textbf{ChatGLM2-based DPFinLLM with DP $\epsilon = 2.0$} &\textbf{0.83636} &\textbf{0.57164} &\textbf{0.83636} &\textbf{0.80454}\\
\hline\hline
\end{tabular}}
\end{center}
\end{table}

\begin{table}[htbp]
\caption{Comparison Performance of Sentiment Analysis on TFNS Dataset (Baselines v.s. Our DPFinLLM)}
\label{tab:tfns_results}
\renewcommand{\arraystretch}{1.3}
\begin{center}
\resizebox{\linewidth}{!}{\begin{tabular}{c||c|c|c|c}
\hline\hline
\multicolumn{1}{c||}{\bf Metric} &\multicolumn{1}{c|}{\bf Accuracy} &\multicolumn{1}{c|}{\bf F1 macro} &\multicolumn{1}{c|}{\bf F1 micro}  &\multicolumn{1}{c}{\bf F1 weighted } \\ 
\hline 
Llama2-7B (Llama2-based) &0.38023 &0.40368 &0.38023 &0.29781 \\
FinGPT-llama2-mt (Llama2-based) &0.78182 &0.67878 &0.78182 &0.75981 \\
FinGPT v3.2 &0.88986 &0.86065 &0.88987 &0.88859\\
\textbf{Llama2-based DPFinLLM with DP $\epsilon = 4.0$}  &\textbf{0.73199} &\textbf{0.60420} &\textbf{0.73199} &\textbf{0.71027} \\
\hline
ChatGLM2-6B (ChatGLM2-based) &0.33124 &0.33976 &0.33124 &0.18787 \\
FinGPT v3.1 (ChatGLM2-based) &0.88275 &0.84917 &0.88275 &0.88214 \\
\textbf{ChatGLM2-based DPFinLLM  with DP $\epsilon = 6.0$} &\textbf{0.72069} &\textbf{0.55256} &\textbf{0.72069} &\textbf{0.68417} \\
\hline\hline
\end{tabular}}
\end{center}
\end{table}

\begin{table}[htbp]
\caption{Comparison Performance of Sentiment Analysis on NWGI Dataset (Baselines v.s. Our DPFinLLM)}
\label{tab:nwgi_results}
\renewcommand{\arraystretch}{1.3}
\begin{center}
\resizebox{\linewidth}{!}{\begin{tabular}{c||c|c|c|c}
\hline\hline
\multicolumn{1}{c||}{\bf Metric} &\multicolumn{1}{c|}{\bf Accuracy} &\multicolumn{1}{c|}{\bf F1 macro} &\multicolumn{1}{c|}{\bf F1 micro}  &\multicolumn{1}{c}{\bf F1 weighted } \\ 
\hline 
Llama2-7B (Llama2-based) &0.56659 &0.52173 &0.56659 &0.48580 \\
FinGPT-llama2-mt (Llama2-based) &0.58636 &0.59366 &0.58636 &0.58039 \\
FinGPT v3.2 (Llama2-based) &0.62762 &0.63837 &0.62762 &0.62728 \\
\textbf{Llama2-based DPFinLLM with DP $\epsilon = 2.0$} &\textbf{0.57129} &\textbf{0.52893} &\textbf{0.57129} &\textbf{0.49360} \\
\hline
ChatGLM2-6B (ChatGLM2-based) &0.56041 &0.48992 &0.56041 &0.44952 \\
FinGPT v3.1 (ChatGLM2-based) &0.64072 &0.64878 &0.64072 &0.64068 \\
\textbf{ChatGLM2-based DPFinLLM  with DP $\epsilon = 8.0$} &\textbf{0.56042} &\textbf{0.48894} &\textbf{0.56042} &\textbf{0.44888} \\
\hline\hline
\end{tabular}}
\end{center}
\end{table}

\begin{table}[htbp]
\caption{Zero-shot Performance of Llama2-based DPFinLLM Fine-tuned on Various Datasets}
\label{tab:zero_shot_llama2}
\renewcommand{\arraystretch}{1.3}
\begin{center}
\resizebox{\linewidth}{!}{\begin{tabular}{c||c|c|c|c|c}
\hline\hline
\multicolumn{1}{c||}{\bf Fine-tuned on:} &\multicolumn{1}{c|}{\bf FPB} &\multicolumn{1}{c|}{\bf FIQA} &\multicolumn{1}{c|}{\bf TFNS}  &\multicolumn{1}{c|}{\bf NWGI}  &\multicolumn{1}{c}{\bf Llama2-7B base model}\\ 
\hline
FPB  &-       &0.40696 &0.56203 &0.40917 &0.40989 \\
FIQA &0.49795 &-       &0.37072 &0.78235 &0.77448  \\
TFNS &0.65201 &0.31464 &-       &0.29436 &0.29781 \\
NWGI &0.62803 &0.47925 &0.46718 &-       &0.48580 \\
\hline\hline
\end{tabular}}
\end{center}
\end{table}

\begin{table}[htbp]
\caption{Zero-shot Performance of ChatGLM2-based DPFinLLM Fine-tuned on Various Datasets}
\label{tab:zero_shot_chatglm2}
\renewcommand{\arraystretch}{1.3}
\begin{center}
\resizebox{\linewidth}{!}{\begin{tabular}{c||c|c|c|c|c}
\hline\hline
\multicolumn{1}{c||}{\bf Fine-tuned on: } &\multicolumn{1}{c|}{\bf FPB} &\multicolumn{1}{c|}{\bf FIQA} &\multicolumn{1}{c|}{\bf TFNS}  &\multicolumn{1}{c|}{\bf NWGI}  &\multicolumn{1}{c}{\bf ChatGLM2-6B base model}\\ 
\hline
FPB  &-       &0.36003 &0.61345 &0.34909 &0.36403 \\
FIQA &0.80120 &-       &0.34792 &0.80343 &0.80312  \\
TFNS &0.18960 &0.18718 &-       &0.18662 &0.18787 \\
NWGI &0.46563 &0.44901 &0.46840 &-       &0.44952 \\
\hline\hline
\end{tabular}}
\end{center}
\end{table}

\begin{figure}[htbp]
  \centering
  \includegraphics[width=\linewidth]{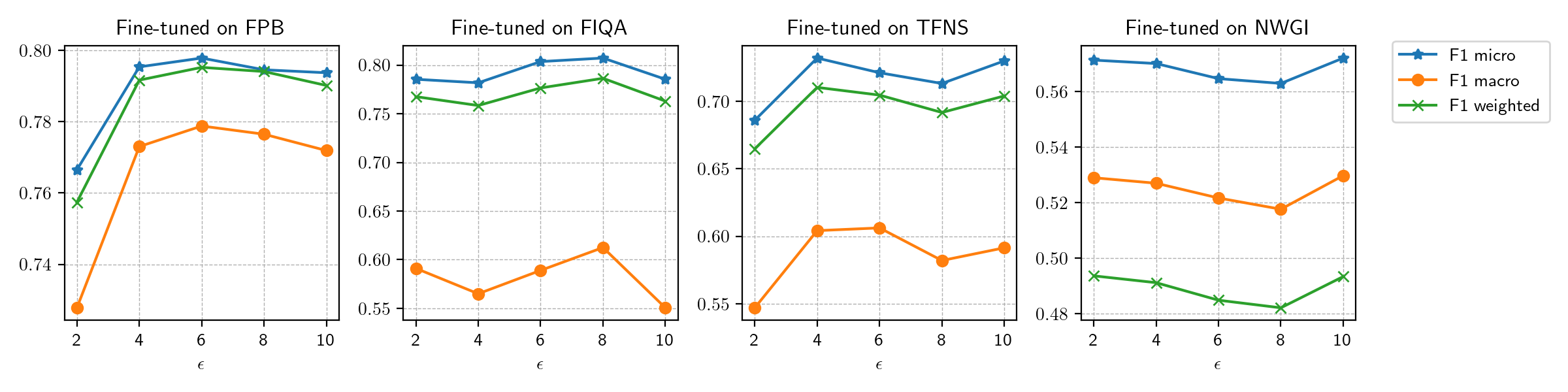}
  \caption{DPFinLLM's Performance with Different Values of $\epsilon$}
  \label{fig:eps_trend}
\end{figure}

\subsection{Evaluation Results on Our DPFinLLM}\label{subsec:cmp_result_alz}

Trained on each dataset, our fine-tuned models—both the Llama2-based DPFinLLM and the ChatGLM2-based DPFinLLM—demonstrate superior performance on test data compared to their respective base models across all evaluation metrics, as shown in Tables~\ref{tab:fpb_results} through~\ref{tab:nwgi_results}. 
By comparing these results in these tables, we draw one conclusion that even under stringent privacy constraints, such as a small $\epsilon$, both the Llama2-based and ChatGLM2-based DPFinLLM models exhibit substantial increases in accuracy and F1 scores on the four datasets. 
This demonstrates that, with appropriately configured hyperparameters, DPFinLLM can achieve performance comparable to fully fine-tuned models while ensuring robust privacy protection for the training data.
Additionally, for the FPB dataset (Table~\ref{tab:fpb_results}) and the TFNS dataset (Table~\ref{tab:tfns_results}), the Llama2-7B-based DPFinLLM model achieves nearly a twofold improvement in each evaluation metric compared to its base model, Llama2-7B. 
Besides, for the FIQA dataset (Table~\ref{tab:fiqa_results}), the configuration “Llama2-based DPFinLLM with $\epsilon = 8.0$” significantly outperforms “FinGPT-llama2-mt,” which was fully fine-tuned on all four datasets.

Moreover, we conduct comprehensive experiments to evaluate the impact of different $\epsilon$ values in the privacy-preserving fine-tuning of our proposed DPFinLLM. In these experiments, $\epsilon$ is the only variable, while all other hyperparameters remain fixed. The results are summarized in Fig.~\ref{fig:eps_trend}.
As shown in Fig.~\ref{fig:eps_trend}, the FPB, TFNS, and FIQA datasets exhibit a similar trend as $\epsilon$ increases: the F1 scores initially rise to a peak before declining, which indicates that our proposed DPFinLLM can protect data privacy while maintaining the performance of sentiment analysis on financial data. 
However, this observation highlights an important insight for hyperparameter tuning in differential privacy—relaxing the privacy constraint (i.e., increasing $\epsilon$) does not always lead to better performance in terms of F1 scores or prediction accuracy.
Interestingly, the NWGI dataset demonstrates a slightly different pattern with larger $\epsilon$ values. A potential explanation for this deviation could be the nature of the dataset: NWGI instructions are generated by ChatGPT, whereas the other three datasets are manually labeled and curated, potentially leading to differences in data characteristics and model behavior.

\subsection{Zero-shot Performance of Our Proposed DPFinLLM}\label{subsec:zero_shot_perf}

After fine-tuning the model on one dataset, we evaluate its zero-shot performance on the remaining three datasets. The weighted F1 scores are presented in Tables~\ref{tab:zero_shot_llama2} and~\ref{tab:zero_shot_chatglm2}, where each row indicates the dataset the model was fine-tuned on, and each column corresponds to the test dataset used to assess model performance.
(1) In Table~\ref{tab:zero_shot_llama2}, with Llama2-7B as the base model, fine-tuning on TFNS leads to a significant increase in F1 score on the FPB test dataset, and vice versa—fine-tuning on FPB improves performance on the TFNS test dataset. Additionally, fine-tuning on FPB enhances the model’s performance on both the TFNS and NWGI test datasets. For the remaining zero-shot experiments, fine-tuning on one dataset generally does not significantly compromise the model’s generalization ability, except in the case where the model was fine-tuned on TFNS and evaluated on FIQA, which resulted in a notable drop in performance.
(2) Similarly, in Table~\ref{tab:zero_shot_chatglm2}, using ChatGLM2-6B as the base model, we observe that fine-tuning on one dataset does not significantly degrade performance on the other datasets, with the exception of fine-tuning on TFNS and testing on FIQA. While the exception case highlights the ongoing challenge of balancing generalization and specialization in zero-shot settings, fine-tuning LLMs with DP still demonstrates great potential in maintaining the model's zero-shot performance on unseen datasets after achieving strong performance on the fine-tuned dataset.

\section{Conclusion}\label{sec:conclusion}

In this paper, we introduce DPFinLLM, a novel privacy-preserving financial large language model designed to address the growing concern of sensitive data leakage in fine-tuning processes. By integrating differential private training mechanism into the fine-tuning pipeline, DPFinLLM ensures robust protection of sensitive financial data while maintaining competitive performance across sentiment analysis tasks. The model’s lightweight architecture, inspired by state-of-the-art small LLM designs, enables efficient deployment on resource-constrained edge devices.
Extensive experiments conducted on four financial sentiment datasets validate the efficacy of DPFinLLM, demonstrating significant improvements over baseline models even under strict privacy constraints, and the experimental results also highlight DPFinLLM’s ability to balance generalization and specialization, achieving superior zero-shot performance across unseen datasets. This capability underscores its potential as a versatile tool for on-device financial applications, where privacy and accuracy are paramount.
To sum up, by combining the differential privacy training idea with an efficient architectural design, the proposed DPFinLLM marks a significant breakthrough in securing financial LLMs, paving the way for the broader adoption of privacy-preserving technologies in on-device financial applications.


\bibliographystyle{IEEEtran}
\bibliography{DPSGD_FINGPT}

\end{document}